 \documentstyle[aps,preprint]{revtex}

\begin{document}
\draft

\title{\bf Spinodal instabilities within BUU approach}

\author{ S. Chattopadhyay}

\address{ S. N. Bose Centre for Basic Sciences, DB-17, Bidhannagore,
Calcutta-700064, India}
\maketitle

\begin{abstract}
Using a recently developed method for the inclusion of fluctuation in the BUU
dynamics, we study the self-consistent propagation of inherent thermal noise
of unstable nuclear matter. The large time behaviour of the evolving system
exhibits synergism between fluctuation and non-linearities in a universal
manner which manifest in the appearance of macroscopic structure in the
average description.
\end{abstract}

\narrowtext
\newpage
In recent years a considerable interest has grown in the understanding
of the role of fluctuation in the dynamics of multifragmentation.
In the decompression phase of the collision, the density
of the composite nuclear system eventually attains such a low value so that
it crosses the boundary of the so called spinodal zone where a small
perturbation
can grow rapidly until  the fragments form. The fluctuation certainly
plays a significant role in the process of fragmentation.
To unravel some basic features of such delicate process
simple model study seems encouraging.
For this
purpose, different theoretical approaches based on Landau \cite{Peth,Kide}
or Boltzmann-Uhling-Ulhenbeck (BUU) \cite{Burg} equation
have been utilized.

 Let us concentrate on the BUU approach to the problem.
In this model, the description of the process is
governed by the evolution of reduced one body density distribution. In the mean
field approximation, the fluctuations arising from the stochastic part of the
collision term are ignored and evolution of a single average effective density
is considered only. In this description the bundle of trajectories are confined
in a
certain region of the associated phase space. The presence of collision term in
the equation corresponds to the dissipative
effect in the evolution. When the system reaches the point of instability
where any fluctuation, however, small it is, amplifies indefinitely so that
trajectories may bifurcate and drive the system to some new fixed point.
In other words, different manifestation of the system are  now become
accessible
dynamically. Therefore, the mean trajectory description has been lost. To
encounter
such a situation, the general strategy is to study the evolution of a
collection
of initially prepared samples. Due to the very presence of the stochasticity in
the
process their fates are different. To make this picture consistent, a usual
method
is to introduce a fluctuation term in the equation of motion, so that the
nature
of the solution become probabilistic. Within the framework of BUU equation
this prescription is properly incorporated and the resultant evolution equation
for phase space density is commonly
known as the Boltzmann-Langevin (BL) equation \cite{Ayik}.

Guided by the same physical principle, several computational schemes have been
proposed to
incorporate fluctuation in the BUU equation \cite{Burg1,Sura}.
In this context, we want to mention
a very recent development (see Ref. \cite{Chat} where a novel method of
simulation is proposed to incorporate fluctuations in the BUU equation. The
 occupancy factor $n({\bf r, p})$ at a particular elementary cell $\Delta {\bf
s}$
of size $\Delta{\bf r}\Delta{\bf p}/h^D$ (D the dimension of the physical
space)
around a point $({\bf r, p})$ in the
associated phase space of the one body description of the system is considered
to
be discrete
in nature so that it takes the value 1 (say success) or 0 (failure) otherwise.
The basic
transition probability (say, $12\to 1'2'$) within a small interval
of time is given by
${\cal T}(12\to 1'2')=\omega_{12\to 1'2'} \Delta t\Delta {\bf s}^4n_1n_2\bar
n_{1'}\bar n_{2'}$
where $\omega_{12\to 1'2'}\Delta {\bf s}^4$ is the basic transition rate. For a
particular
transition, every sample is tried with a probability of success $\omega \Delta
t\Delta{\bf s}^4$.
However due to the presence of four occupancy factors in the above relation
a successful transition  occur when an appropriate configuration of particle
and
hole states appeared in a sample. By this way enough fluctuation in sample
space
can be generated and at the same time, the relevant statistical criteria i.e.
$\sigma^2_i=\langle n\rangle_i\langle\bar n\rangle_i$ can be
fulfilled exactly for every cell i.

  Another essential aspect of the simulation is to generate the solution of the
Vlasov part of the BUU equation in the presence of the mean field potential.
Due to the
discrete nature of the phase space
occupancy $n({\bf r, p})$, the usual procedure of matrix method
\cite{Burg,Burg1} cannot be
applied here properly. To consider Vlasov propagation of phase space
distribution
we follow the Nordheim's approach, closely related to the well known test
particle
method of solving Vlasov equation. A filled unit phase cell may be considered
as a
particle. The time evolution of these particles can be described by the
standard
leap-frog algorithm in the following manner,
\begin{eqnarray}
X^i_k(t+\Delta t)&=&X^i_k(t)+{1\over m}P^i_k \Delta t~;\\
P^i_k(t+\Delta t)&=&P^i_k(t)+\nabla_k{\cal U}(\rho({\bf r},t))\Delta t
\end{eqnarray}
where ($X^i_k(t),P^i_k(t)$) represent the $k$th component of position and
momentum of
the ith particle at time $t$. It is important to note that unlike to the
test particle method the position  and also momentum of the particle are
represented by the grid points here, therefore, its evolution with time take
places
in discrete steps. However to provide a reasonably good
energy conserving evolution, we incorporate a modification in the leap-frog
routine. The error comes mainly due to the rounding off  the quantities
$X^i_k/x_d$
or $P^i_k/p_d$ where $x_d$ and $p_d$ are the co-ordinate and momentum grid
spacing.
The accumulated error that arise in momentum due to the incoming particles at
a particular spatial cell (say $l$) is given as
$\Delta\tilde p^l_k=\sum_i\delta p^i_k$.
Here $\delta p^i_k$ is
the residual part of $P^i_k$ so that
$P^i_k=p_d\times\hbox{nearest integer of} (P^i_k/p_d)+ \delta\tilde p^i_k$.
Therefore,
to reduce this error substantially, one may assume that the momentum of all the
particles in
a cell $l$ are modified  as $p_d\times\hbox{nearest integer of}({P^i_k\over
p_d})+{\Delta\tilde p^l_k\over N_l}$.
Here $N_l$
is the total number of incoming particles in cell $l$.
To check the reliability of this method we consider the evolution of the
breathing
mode of a spherical nucleus in 2-dimension having small skin of about 2 fm.
We observe  two interesting points: 1) the initial spherical symmetry in
momentum
and co-ordinate space are maintained exactly and 2) no two particles appear
simultaneously in the same phase cell throughout the evolution. However, it is
to
be noted that when fluctuation in phase cell is taken into account by allowing
the
the collisions among the particles the above mentioned features are gradually
lost. In this situation, the smearing of the distribution becomes necessary.
This is
incorporated by simply moving a particle in the neighbouring sites in momentum
space
if it tries to access a cell in the process of evolution which is already
filled.
In this simulation, the values
of $x_d$ and $p_d$ are  taken  as 1 fm. and 40$\sim$60 MeV/c respectively.
The total energy per particle is found to be conserved up to 95$\%$ even at
time
$t=100$ fm/c.

Using the above mentioned simulation procedure
the present study revisits the problem of nuclear matter in
spinodal zone. This work mainly concerned about the large time behaviour of the
system and the non-linearities that appear in the process of evolution. To make
this study consistent with earlier investigation we adopt the scenario that was
studied rigorously by several authors \cite{Burg,Burg1} namely, a gas of
fermions situated
inside a two dimensional torus. For the effective one body field at two
dimensional
grid point (x,y) we employ a simplified Skyrme interaction
$U(x,y)=A {\rho(x,y)\over \rho_0}+B({\rho(x,y)\over \rho_0})^2$ with $A=-100.3$
MeV, $B=48$ MeV, the saturation density $\rho_0=0.55$ fm$^{-2}$.
To evaluate effective potential on the lattice an averaging over 2-dimensional
Gaussian function having width 0.87 is done. The in-medium cross section
is taken as 2.4 fm and Fermi-momentum $P_f=260$ MeV/c. For simulation we
consider the square lattice of size (21$\times$21) in co-ordinate space having
width
$x_d=1$ fm., (31$\times$31) in momentum space with $p_d=40$ MeV/c and the
number of
samples ${\cal N}_s$ for parallel runs are taken to be 100. To ensure that our
system
is initially situated inside the spinodal zone we prepare the samples having
uniform (in space) average density $\langle\rho\rangle$ (or the number density
$\langle N\rangle$)=$0.5\rho_0$ $(N_0)$ and momentum distribution
same as that of a fermi gas at temperature $T=3$ MeV.

At a given temperature $T\not=0$ there exists
thermal fluctuation in the initial state.
To incorporate this important feature within our simulation
procedure we prepare the samples with the assumption that the distribution of
$N$ is given by a Gaussian function with mean $\langle N\rangle$ where as the
sample average value of
occupancy $\langle n\rangle_i$ is given by the fermi distribution of
appropriate
temperature $T$. The variance $\sigma^2_N$ satisfies the standard relation
$\sigma^2_N=\sum_N\sum\sigma^2_i=\sum_N\sum\langle n\rangle_i\langle\bar
n\rangle_i$. To maintain spatial
uniformity in initial state, the value of $N$ or the momentum distribution is
considered
to be the same at every spatial cell of a particular sample. Inspite of the
fact that at $T=0$ all collisions are Pauli blocked,
the symmetry of the system remains intact even
by BL treatment used here or the same used in Ref. [3]. It is to be noted that
in this case $\sigma^2_N=0$ i.e. all samples are identical.
The presence of the thermal fluctuation in the initial state is in fact, the
source
of disturbance which generates irregularities subsequently in co-ordinate
space and may amplify due to the action of the mean field. Hence, to make the
picture consistent, the preparation of the samples in the aforesaid manner is
therefore an essential part of our simulation procedure.

Let us now turn to the investigation of the Boltzmann-Langevin dynamics on
lattice. The normal mode analysis of the evolving density will allow us to
study the interplay between instabilities and fluctuations. Following the
Ref.\cite{Burg} we introduce two point correlation function to the associated
density fluctuation as
\begin{equation}
\sigma_{\bf k}(t)=\langle|\delta\rho_{\bf k}(t)|^2\rangle={1\over
L^4}\int^{L^2}d{\bf r}'
\int^{L^2}d{\bf r}e^{-{\bf k}.|{\bf r}'-{\bf r}|}\langle\delta\rho({\bf r}')
\delta\rho({\bf r})\rangle
\end{equation}
For theoretical analysis of the time evolution of the quantity $\sigma_k$ the
starting point is to linearise the Vlasov
equation and derive the equation for normal modes (for example see Ref.
\cite{Colo}).
In the presence of fluctuation in the transport process the evolution of
the amplitude of the Fourier components (or in other words, that of the
collective
mode $\nu$)
can be represented by the following Langevin equation
\begin{equation}
{{dA_\nu(t)}\over{dt}}=i\omega_\nu A_\nu(t)+\tilde{\cal B}_\nu(t)
\end{equation}
where $A_\nu$ is the two component amplitude matrix (with  elements
$A^+_\nu$ and $A^-_\nu$) of the Fourier components of the density fluctuation
$\delta f_\nu$ so that $\delta f_\nu=A^+_\nu e^{i\omega_\nu t}+A^-_\nu
e^{-i\omega_\nu t}$.
$\tilde{\cal B}_\nu(t)$ is the kicking term of the Langevin equation. The
associated correlation or diffusion matrix for Gaussian Markov process can be
given by a Hermition matrix with components
\begin{equation}
\langle{\cal B}^i_\nu(t){\cal B}^j_\nu(t')\rangle={\cal D}^{ij}_\nu\delta(t-t')
\end{equation}
where i (j) runs for states '+' or '-'.

Let us now concentrate on the situation of instability. For infinite system
one can find a range of $\nu$ , for which all the modes are unstable,
the associated
the frequencies $\omega_\nu(=-i\Omega_\nu )$ are purely an imaginary numbers
and beyond
that range all modes are stable. Therefore, in course of time the fate
of the system is dictated by the evolution of few such most unstable modes.
The probabilistic evolution of $A_\mu$ (see eq.(4)) can be described as well in
terms of Fokker-Planck equation. The evolution equations for first and second
moment related
to it can be written as \cite{Risk}
\begin{eqnarray}
{dA_\mu\over{dt}}&=&\pmatrix{\Omega_\nu &0\cr 0&-\Omega_\nu }\langle
A_\mu\rangle(t)\\
{d\sigma^{ij}_\nu(t)i\over {dt}}&=&\Omega^{il}_\nu\sigma^{lj}_\nu(t)
+\Omega^{jm}_\nu\sigma^{mi}_\nu(t)+2{\cal D}^{ij}_\nu
\end{eqnarray}
where $\sigma^{ij}_\nu$ is a symmetric matrix.
The initial condition i.e at t$=0$, $d\sigma_\nu /dt=0$
($\sigma_\nu=\sum_{i>j}\sigma^{ij}_\nu)$
satisfies if ${\cal D}^{++}_\nu={\cal D}^{--}_\nu=-{\cal D}^{+-}_\nu$.
These equations are the same as those derived in
Ref \cite{Colo}, the solutions of which provide  a good fit to the
simulation result \cite{Burg2}.

Within the framework of Langevin or FP equation, further study can be directed
to the non-linear regime of the dynamics by introducing
non-linearity in the drift term.
Since $-dU_\nu/dA_\nu$ correspond the drift coefficient of FP equation
this can be done by adding a quartic term to the associated quadratic
potential.
This stochastic model of non-linear Langevin equation, is  rich in physics
capable of explaining the general behaviour of the system in the presence of
instability far from the equilibrium and onset of macroscopic structure
\cite{Suzu}.
The presence of the quartic term prevents an indefinite fall of a system and
drive it towards any one of the two newly appeared stable fixed points
which correspond to the minima of the dynamically generated cusps.
The co-operative behaviour or the synergism of nonlinearity
and random force simulate this typical feature i.e. {\it slowing down} the
fluctuation near dynamical
phase transition. The fluctuation deviate
gradually from its initial Gaussian profile as the system approaches to the
second or scaling regime. The modified equation for second moment
with positive value of $\Omega$ is given by
\begin{equation}
{d\over{dt}}\sigma^{++}_\nu(t)=2\big\{\Omega_\nu-g_\nu\sigma^{++}_\nu(t)\big\}
\sigma^{++}_\nu+2{\cal D}^{++}_\nu
\end{equation}
which provides a scaling solution
\begin{equation}
\sigma^{++}_\nu(t)={\Omega_\nu\over g_\nu}{\tau\over{1+\tau}}~;~~~~
\tau={g_\nu\over\Omega_\nu}\big(\sigma^{++}_\nu(0)+
{{\cal D}^{++}_\nu\over\Omega_\nu}\big)e^{2\Omega_\nu t}
\end{equation}
with $g_\nu>0$.
However, in the
derivation of Eq. (8) a physical approximation i.e.
$A^{++3}_\nu(t)=\sigma^{++}_\nu(t) A^{++}_\nu(t)$ is made \cite{Suzu}. The time
needed to reach the saturation
for a unstable mode $\nu$ is given by
\begin{equation}
t^0_\nu={1\over{2\Omega_\nu}}log\big[{g_\nu\over\Omega_\nu}\{\sigma^{++}_\nu(0)+
{{\cal D}^{++}_\nu\over\Omega_\nu}\}\big]^{-1}
\end{equation}

In Fig. 1. the open diamonds represents the time evolution of $\sigma_{\bf k}$
for
unstable modes with values of $k_x$ ranging from 0.3 to 0.6 with $k_y=0$.
For direct comparison to earlier results \cite{Burg} we
plot the function $L^2\sigma_{\bf k}$ (see Fig. 1.) so that the unit becomes
fm$^{-2}$.
The solution of Eqs. (7) (valid for linear regime of the dynamics)
are shown by dotted curves. The solid curves represent the solution of
non-linear
generalization  so that the evolution of $++$ component is given by Eq. (9).
For these solutions we use the initial condition i.e. at $t=0$, $\langle
A\rangle_\nu=\sigma_\nu=
d\sigma_\nu /dt=0$.
The general agreement of our fit to the simulation
result is quite good. The saturation time $t_{\bf k}^0$  for
different modes are shown in the figure.  The most unstable mode
i.e. the mode having minimum growth constant $t_{\bf k}$ ($=\Omega^{-1}_{\bf
k}$)
saturate earliest and the time $t_{\bf k}^0$ gradualy increses as the $t_{\bf
k}$ increases.
For some cases, the linear regime of the dynamics persists even at time $t=120$
fm/c. Let us consider the evolution of density $\rho(x,y;t)$ in co-ordinate
space. In earlier time up to $\approx 75$fm/c when spatial average of
$\sqrt\langle(\delta\rho)^2\rangle/\langle\rho\rangle$ reaches the value 1 we
observe a considerable
fluctuation
in $\rho$ in the individual samples. However, the sample average density i.e.
$\langle\rho(x,y;t)\rangle$ is found to be almost uniform with $\sim 3\%$
fluctuation
from its initial value of 0.5 which
is essentially the mean (or single) trajectory result valid in the linear
regime of the dynamics.
As the time increases the non-linear effect makes the situation different. We
see structure even in the sample average distribution of density which
develops slowly in a coherent manner. In Fig. 2 we plot
$\langle\rho(x,y;t)\rangle$
at a time $t=150$fm/c. This observed structure almost remains steady with
increasing
time, which
corresponds to a equilibrium or at least a metastable state.
The fluctuation is seen to be $\sim 20\%$ with respect to the initial
uniform state.  The appearance of
such spatial inhomogeneity in the average density is in conformation
with that of a mean
trajectory calculation reported in Ref. \cite{More}.  Though,
the fluctuation in latter case is larger than what we observe here.
This may be attributed to the different initial and also boundary
conditions used in these two cases.
However,
large scale fluctuation exists in individual samples, as evident from
the attainment of a non-zero steady value of $\sigma_{\bf k}$ in the
present formalism.

In conclusion, using a recently proposed simulation scheme for inclusion
of fluctuation in BUU dynamics we study the problem of spinodal decomposition
in nuclear matter. Our work is mainly devoted to the study of the large time
behaviour of system. To analyize the simulation result we apply the
well known Suzuki's model of dynamical phase transition successfully
in the context of spinodal decomposition. Accordingly,
the evolution of the system is guided by the co-operative effect of
non-linearity
and fluctuation in a universal manner. To make our stand clear, we would
like to
mention a recent work done by Baldo et al. \cite{Bald} where the evolution
of fluctuation is studied in the framework of Vlasov equation.  In the
simulation
work they studied the growth of an initial disturbance imposed on the uniform
nuclear matter. Without being saturated, the $\sigma_{\bf k}$ overshoots very
fast from its linear trends (in log scale). It will be relevant here to note
that the finite collision rate, no matter how small it be,
changes the nature of the solutions describing a
diffusive process as modeled by the Langevin equation.
Therefore, the non-diffusive behaviour of fluctuation, although exhibits some
non-standard
evolution pattern, but does not indicate conclusively that the dynamics of
fragmentation is dominantly non-linear.
However, similar such studies facilitate us to extract the growth constant
$t_{\bf k}$,
(the RPA frequency) of
the initial state \cite{Burg2} and also allow a further scope to
check the consistency of our work. The values of $t_{\bf k}$ indicated in Fig.
1.
provide a good fit to the dispersion relation shown in Fig 3. of Ref.
\cite{Burg2}.

\newpage
\centerline{{\bf Figure Captions}}

\newcounter{bean}
\begin{list}%
{Fig.\arabic{bean}}{\usecounter{bean}}
\item  The fluctuation coefficient $L^2\sigma_{\bf k}$ versus time $t$ for
different values of unstable mode ${\bf k}$ ($k_x,k_y$) are shown. The
open diamonds correspond to simulation results in steps of 10 fm/c. The dashed
curves represent
the solution of Eq.(7) while solid curves show the nonlinear generalization
of the same. The open circles represent the simulation result of a typical
stable mode with ($k_x= 0.8$ fm$^{-1}$, $k_y=0$) which exhibits an early
saturation.

\item  The sample average density profile $<\rho(x,y;t)>$ at time $t$= 150 fm/c
versus
the position ($x$, $y$) is shown.

\end{list}
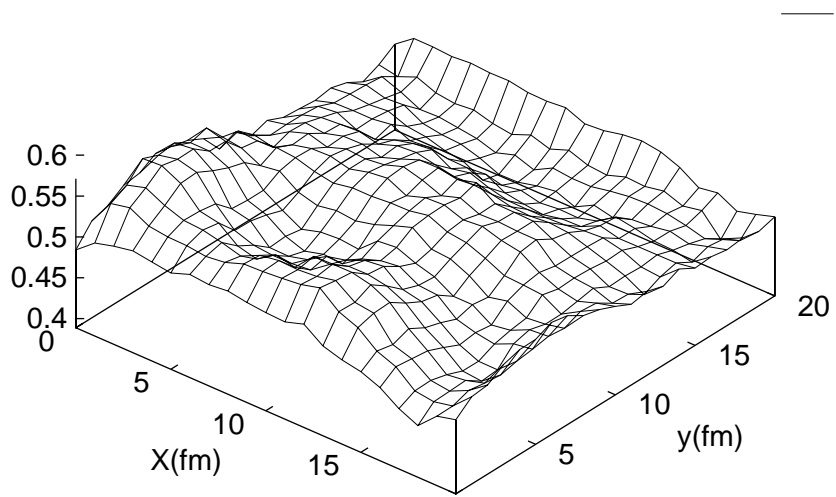
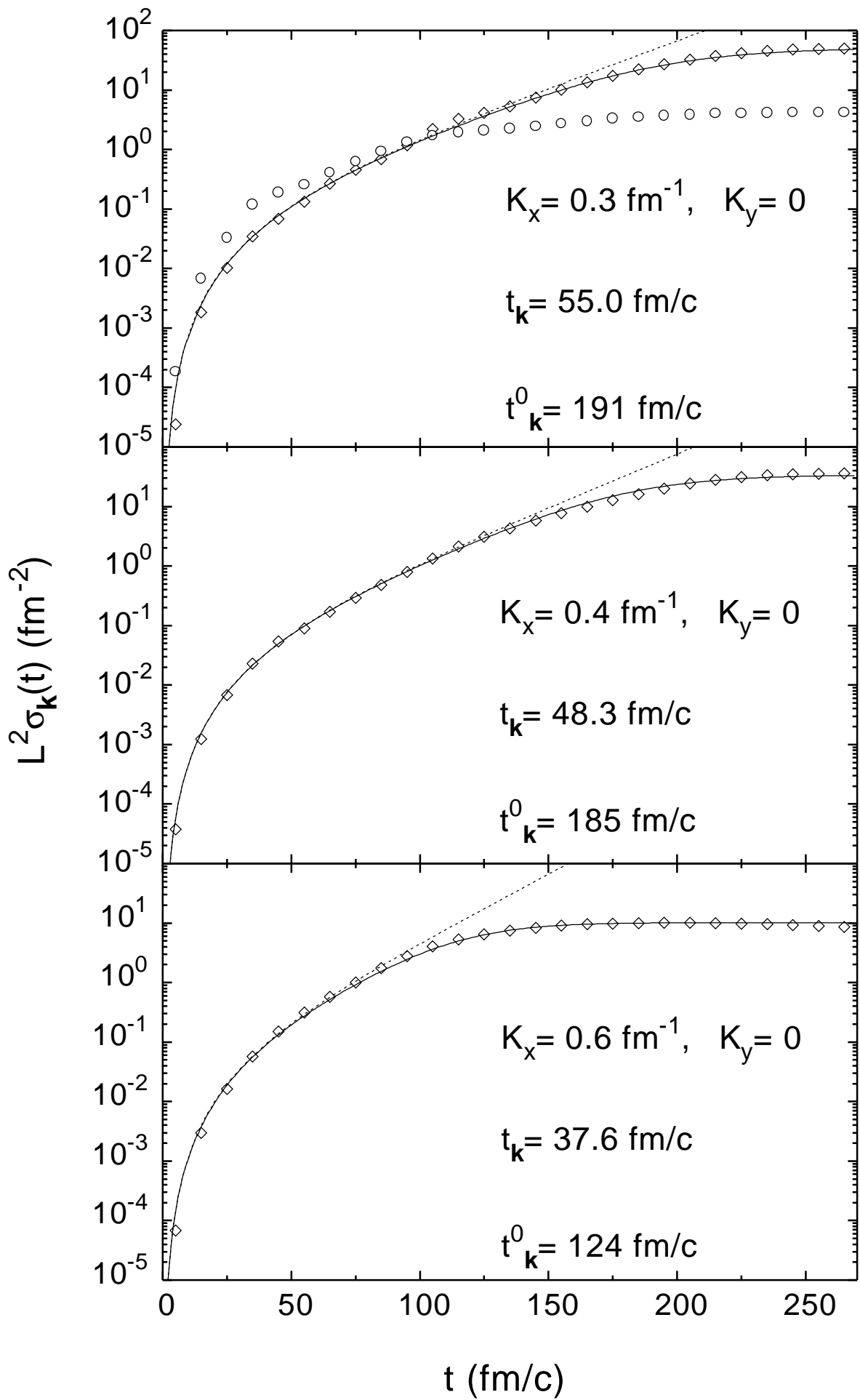
\end{document}